# Application of Blockchain Frameworks for Decentralized Identity and Access Management of IoT Devices

Sushil Khairnar
Virginia Tech, Blacksburg, 24060-61, USA

*Abstract*—The growth in IoT devices means an ongoing risk of data vulnerability. The transition from centralized ecosystems to decentralized ecosystems is of paramount importance due to security, privacy, and data use concerns. Since the majority of IoT devices will be used by consumers in peer-to-peer applications, a centralized approach raises many issues of trust related to privacy, control, and censorship. Identity and access management lies at the heart of any user-facing system. Blockchain technologies can be leveraged to augment user authority, transparency, and decentralization. This study proposes a decentralized identity management framework for IoT environments using Hyperledger Fabric and Decentralized Identifiers (DIDs). The system was simulated using Node-RED to model IoT data streams, and key functionalities including device onboarding, authentication, and secure asset querying were successfully implemented. Results demonstrated improved data integrity, transparency, and user control, with reduced reliance on centralized authorities. These findings validate the practicality of blockchain-based identity management in enhancing the security and trustworthiness of IoT infrastructures.

## I. Introduction

Internet of Things (IoT) is an emerging field of research which is about connecting smart devices, people and systems together using sensors, software and the other technologies to capture meaningful information from the surrounding environment of devices and take actions to improve productivity and efficiency. There is an increase demand for IoT devices ranging from Google Home and Amazon Alexa to Smoke Alarm and the next Door bell. In 2021, there were more than 10 billion active IoT devices. By 2025 it is estimated that there will be over 150,000 IoT devices connecting to the internet per minute. It is predicted that the number of active IoT devices will surpass 25.4 billion by 2030 [2][17]. However, the proliferation of IoT devices all over the world raises concerns over security, privacy and device management. IoT devices generate huge amounts of data most of which is sensitive data belonging to the individuals and the companies. One of the most popular IoT based attacks was the Distributed Denial of Service (DDoS) attack of 2016 which caused major Internet platforms and services to be unavailable to large swathes of users in Europe and North America [3][28].

In the case of James Bates of Arkansas, USA, who was charged with first-degree murder partly with the help of evidence collected by an Amazon Echo smart speaker in 2015 [4][5][6]. The prosecutor used the data found on Bates Amazon Alexa data along with his smart meter data to build the case. During the case, Amazon refused to release the data collected using Alexa. However, the defendant gave permission to use the data during the case. According to the reports, the Echo wasn't the only smart device investigators cited in their case - they also used "information from a smart water meter, alleging that an increase in water use in the middle of the night suggests a possible cleanup around the crime scene"[23]. The case was dismissed in December 2017, but not before becoming a news item and drawing the defendant's personal life into the press[24] [11].

IoT finds its applications in various areas including health care, smart homes, self driven cars and smart cities. The data captured from the devices is used to gain insights and make decisions. This decision making process "trusts" in the information obtained. There- fore, "trust" is the pillar of the advancements in IoT. To ensure that the captured data is obtained from reliable sources, blockchain technology plays a crucial role. Currently, more than 10 billion de- vices are connected over internet through IoT. Inorder to ensure the "trust" we must be able to verify and validate the identity of these devices through out their life cycle. The identity of these devices is tracked by blockchain [8]. Since decentralization is at the heart of the blockchain technology, we leverage this to eliminate centralized identity management of devices[1].

The device management particularly, is considered a critical task in IoT applications because of the large number of devices connected and the multifariousness of the network. Decentralized identifier (DID) is a new model for creating global unique identifiers. These identifiers are built on top of distributed ledgers. DIDs are created as the foundation for decentralized digital identity and public key infras- tructure (PKI) for the Internet. Among the main features is that DIDs offer full control of identity to consumers without any dependence on the central authority. DIDs are defined by a generic format URL that is used to address DID documents and to be resolved by a DID resolver. DID generic format is a URI scheme conformant with the web standards as per RFC3986 [25]. A DID URL is a network loca- tion identifier for a specific resource. It can be used to retrieve things like representations of DID subjects, verification methods, services, specific parts of a DID document, or other resources [26]. The things included in DID URL include the urn scheme -'did' in this case, followed by the namespace or method name (e.g. ethr, sov, etc), and after that is the method-specific identifier which can be a combina- tion of letters and numbers (e.g."did:example:1234AbxfgHUfgh").

However, because of the decentralized nature of blockchain, there is a severe limitation over its performance. For example, Bitcoin can only achieve a low throughput of 7 transactions per second (TPS), and takes over 10 minutes for a transaction to get confirmed [7]. Contrary to the current centralized payment systems such as Master- card and Fedwire can reach thousands of transactions per seconds and almost real-time payments [14]. Performance evaluation and analysis plays an important role to find bottlenecks in the blockchain systems and can be used to inspire further optimization ideas. The performance indicators of the blockchain mainly include transaction throughput and latency. Transaction throughput represents the num- ber of transactions that can be processed at a fixed time, and latency represents the response and processing time to transactions [27]. In addition to these two indicators, resource utilization and ease of use can also be considered among the metrics for blockchain systems evaluation.

This research addresses the question: Can blockchain-based decentralized identity management systems enhance trust, authentication, and ownership control in dynamic IoT environments while maintaining data integrity and privacy?

The remainder of this study is organized as follows: Section II provides an overview of the system architecture. Section III details the design principles. Section IV outlines the implementation steps. Section V evaluates the performance. Section VI compares related work, followed by a new discussion in Section VII. We conclude and present future work in Sections VIII and IX.

## II. OVERVIEW

To bring real trust to the data captured by devices we leverage the core principles of blockchain. The blockchain is a digital ledger of past transactions and has found massive use cases in areas other than cryptocurrency. Its strengths make it an ideal component for IoT solutions. Its design eliminates the need for centralized authorities to govern the system, the transactions stored in every block of the blockchain are completely auditable and are public, these transactions are verified and validated through consesus of the peers and the immutability of the blockchain makes the system tamper proof and trustworthy[13]. We develop on these principle to provide unique identities to devices at the time of their creation. Every device is in control of its identity. This device publishes its public key to the distributed ledger and gets a DID (Document Identifier) in response. Every device presents this DID to the blockchain it wishes to add a transaction to. The blockchain then verifies its identity using this DID and validates it [7]. The digital identity for users works on the above principle. We applied this mechanism for IoT devices. As a blockchain is write forward and immutable, blockchain-born devices thus will have irreversible reputations and identities. We also examined the possibility of creating unique identifiers that are tamper-resistant by combining a sensors environmental inputs like precipitation, temperature, humidity and more. For access management we use smart contracts between parties involved.

We used Node-RED to load and test our blockchain systems over hyperledger framework. Node-RED is a programming language for connecting hardware, APIs, and web services. It is primarily a visual tool for the Internet of Things built by the IBM Emerging Technology organization and is an open source.

Following are some of the most important features of Node-RED architecture which led us to select this tool:

- Node-RED (latest version 0.16) is quick since it runs on the most recent version of Node.js that is supported (LTS 6.x, 6.x).
- A single-threaded event queue that simplifies things.
- Event-driven architecture and asynchronous I/O
- The entire architecture was created with express, d3, jquery, and ws.
- It enables a lightweight runtime environment, as well as an event-driven and non-blocking approach using Javascript
- JSON is used to hold the numerous flows made in Node-RED, which can be readily imported and exported for sharing with others.

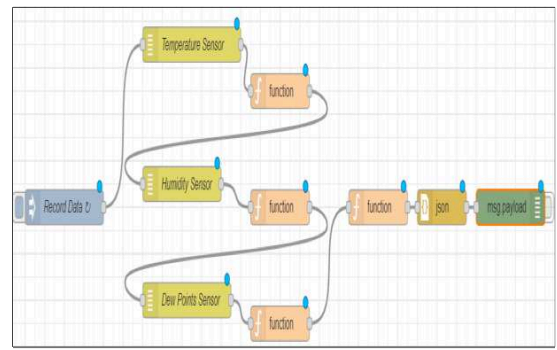

Fig. 1. IoT Network simulation using node-red.

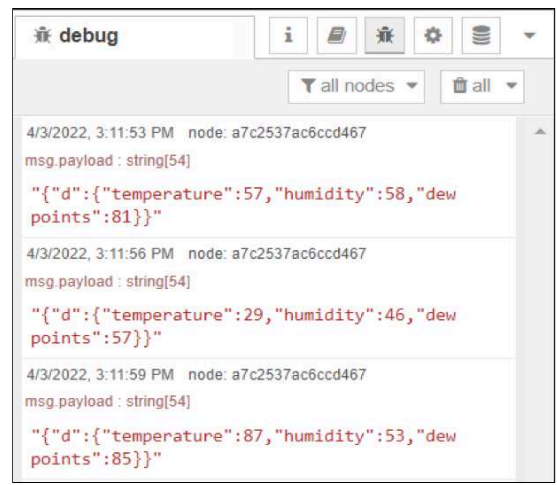

Fig. 2. Node-red - message payload.

In order to understand the usage of Node-RED, simulate an IoT network, and gather transactional data in it, we replicated a small network consisting of three sensory nodes. To generate the data, random nodes are used as sensors which output values in a given range. Timestamp node is used to record data every three seconds. The generated data flows as transactions through the network and accumulated as json output at the end of network. Fig. 1 shows how this network is simulated on Node-RED. Fig. 2 shows the aggregated output data in the json format.

We covered the following cases for device access management:

- Registering a new IOT device in the Blockchain
- Login feature for the registered IOT devices
- Viewing all existing assets or transactions in the blockchain system.
- Viewing assets owned by the device that is currently logged in.
- Uploading a new IOT device data to the blockchain.

## III. Design

A device's life cycle in IoT networks begins with its manufacture and ends with its disposal. The suggested decentralized IdM architecture displays a cycle that specifies the status of the device and who owns it. The device identity must be maintained throughout this life cycle through a global registry that allows authorized registrants (for example, manufacturers) to create new identities for new devices. It enables the registrant to carry out identity management duties or transfer ownership of the device(s) to other parties, such as operators and end-users.

The following are some of the issues we plan to address in regard to the IAM module of the IoT management framework:

- Establish a global and standardized register for IoT devices.
- As the lifecycle of IoT devices begins, define explicit registration processes.
- Ensure that security measures are implemented during the registration process, as compromising the IoT device's identification threatens the sensitive data managed by the IoT device.
- Create an identity lifecycle that can be applied to items in an organization and customized based on the device's lifetime.
- Provide tools for managing device ownership and transferring ownership.
- Provide a way for tracing the device's lifecycle back to its source.
- The construction of a device's digital identity and the necessity for identity formation to reflect device lifecycle changes [30].

The identity creation is performed when the owner submits a request to the IdM smart contract to produce a new DID, which is carried out according to the protocol outlined below. When the contract succeeds, it commits the DID record to the blockchain and stores the DID Document on the hub [29] stated in Algorithm I.

---

**Algorithm 1** Identity Creation
**Input:** publicKey, owner, id, hash, owner
**Result:** (DID) Committed on Chain
**Function** DIDGenerator(publicKey, id, owner, proof);
 Initialization();

**if** *Sender is Authorized* **then**
  CreateDIDIdentifier(id)  **if** *DID doesn't exist* **then**
   valid ← VerifyProof(publicKey,proof)  **if** *valid* **then**
    hash ← storeIPFS(DIDDocument)  ListDID[DID]
    ← hash
   **else**
    pass;
   **end**
  **else**
   **Return** (Identity already exists);
  **end**
**else**
  pass;
**end**

---

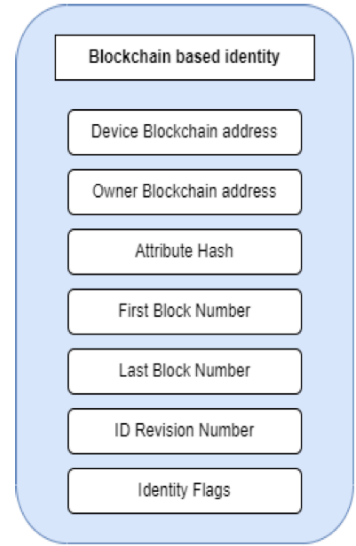

Fig. 3. Digital identity data structure.

The above-mentioned device identity ownership management needs are addressed in this study using blockchain and smart contracts. The immutability and provenance capabilities of blockchain, as well as the digital identity presented in the preceding section, are used to ensure identity ownership to meet the requirements. The identity data structure is modeled using decentralized identifiers (see Fig. 3), enabling secure and user-centric device registration [9].

In addition to the other features, the blockchain identity generated at the moment of device birth is built with a set of attributes that includes the blockchain addresses of the device itself and the original owner (maker). The produced identity is kept in the distributed ledger with a number of properties, the most essential of which is the owner address in this case. The owner's blockchain address is used to identify ownership, and this feature is only accessible to the owner thanks to smart contract structures. As a result, only the device identity owner can transfer ownership, ensuring a secure transfer of ownership from one entity to another. Furthermore, the change of ownership is predicated on the former owner being replaced.

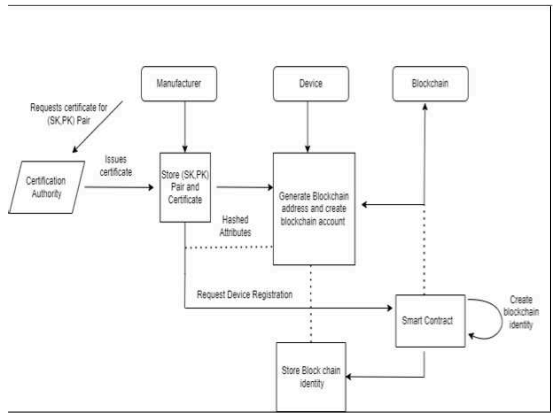

Fig. 4. Device identity management architecture.

Public-key cryptography and hashing functions are used to create identities (Fig. 4). The identity is made up of the minimum set of attributes, that are required to provide a unique and global identity for the device in the global registry. This is accomplished in two steps: the creation of cryptographic keys and blockchain addresses, as well as the creation of a blockchain-based digital identity. The decentralized identity technique suggests utilizing cryptographic keys to create anonymous IDs or using the device technology unique identifier (e.g., IMEI or UUID) as the DID identification.

All of the electronic devices which are capable of connecting to the internet can act as entities that have ability to be a part of an IoT network. Formation of an IoT network could be impacted by number of factors such as types of the devices, roles these devices play, and possible values of physical distance between the devices etc. However, depending on the functionality, the nodes in the network can broadly be classified into three main components: coordinator, router and the end devices [27]. Some nodes deliver information and other collect and deliver it. Also, a set of nodes control other set of nodes. Formation and simulation of such network and transfer of data from the network to a blockchain will be addressed by using framework known as Node-RED [25].

To carry out these operations smoothly, each entity and device need to be well identified and documented. Identification is not only limited to the devices but it is required for entities such as the users, data, services, etc. Also, one entity can be a part of multiple IoT networks. Hence, depending upon the scope of different networks, assigning unique identity plays key role in managing and ensuring faultless operation of the network. Number of approaches are available for the allocation of new identities. For instance, a common repository can be maintained to allocate unique keys [28]. Apart from this, an algorithm can be used to generate random keys from a sizable pool of keys. If there is hierarchy of networks, separate pool of identities can be maintained to ensure appropriate allocation of identities.

For the implementation of this approach we have divided the system into two components. The first component will be the implementation of device identity management and the second will deal with the application of blockchain in IoT. The former helps in verifying the identity of the device before it is added to the network. In case of the latter, we propose to bootstrap a Hyperledger fabric that resembles a network of temperature sensors in a warehouse. The network consists of organizations and sensors that deploy the chaincode (smartcontract) and drive the execution of the transactions [26].

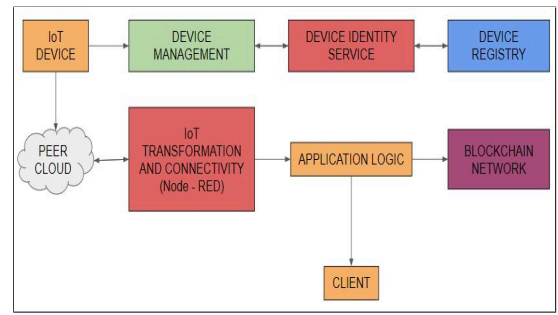

Fig. 5. System architecture.

The components of the system architecture (Fig. 5) are described as below:

*1) IoT device:* This can be a sensor, actuator that has a network connection and may also have a user interface.

*2) Device management:* It provides a way to securely manage devices that become a part of the network.

*3) Device identity service:* It reliably verifies the identity of the device once the device is registered.

*4) Device registry:* It stores the information about the devices.

*5) Peer cloud:* This is a broker into which the measured data from the sensors is pushed at regular intervals.

*6) IoT Transformation and Connectivity (Node-RED):* This is an IoT connector tool by Node.js. Node-RED is a programming tool for wiring together hardware devices, APIs and online services in new and interesting ways [25].

*7) Application logic:* This is the core component of application. It coordinates the handling of data received from Node-RED and passes it to the blockchain network.

*8) Blockchain network:* It is a network of the blockchain nodes that maintain the distributed ledger.

*9) Client:* This is the end-user application.

The runtime flow of the system begins with verifying the identity of the IoT device and the IoT device sending events such as temperature, humidity, etc to the peer cloud along with the timestamp. The IoT transformation and connectivity service performs some event filtering at this step before this data flows into the core component of the application. The application then prepares and maps the data to be able to send it as transactions to the blockchain. Notifications from the blockchain such as settlement of a transaction or violation of a contract are sent to the application. This blockchain network is implemented using the HyperLedger Fabric Model (Fig. 6). Every peer node in Fig. 6 runs in docker containers and relies heavily on smart contracts. Chaincode specifies the asset's business logic, or the rules for reading and changing the asset's so-called state. Fabric, unlike public cryptocurrency blockchains, allows parties to create a private route for their assets, isolating and segregating transactions and a ledger. The end-user application then displays these notifications to the client. This gives the ability to the customers to check the status of their assets by tracking the record by using an application UI. The solution searches the blockchain's transaction history and sends out various forms of notifications on transaction status.

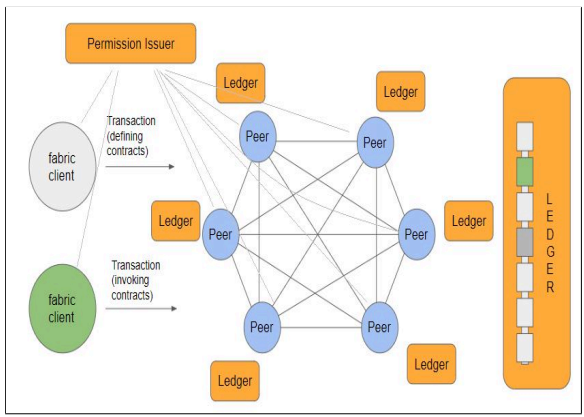

Fig. 6. Hyperledger fabric model.

Hyperledger Fabric is a permissioned distributed ledger framework for building enterprise-grade systems and apps. Its modular and adaptable architecture caters to a wide range of industry applications. It takes a novel method to consensus that allows for scalability while maintaining anonymity. Public blockchains, such as Bitcoin, do not require permission. Anyone is welcome to participate. Hyperlegder, on the other hand, is a permissioned blockchain platform, which means that only validated identities can join.

Some of the common issues with privacy are:

- Scalability: Because all nodes on the network must execute transactions, most public blockchains have low scalability. As a result, transaction throughput is low [10].

- Storage issues: Because every full node on the public blockchain contains all of the data, storage requirements are quite high and are steadily increasing over time. This level of storage redundancy isn't viable for most applications.

- The 'Proof Of Work' (POW): technique, which needs a lot of computational power and is energy-intensive, is used by Bitcoin and Ethereum. As processing power and energy requirements rise over time, it becomes impractical in a corporate setting. Hyperledger Ordering Node, on the other hand, employs a variety of techniques, including Solo, Kafka, and PBFT.

- Lack of governance: Because no one has control over public blockchains, it is up to individual developers or developer communities to implement improvements. To run blockchain properly, businesses require adequate governance [26].

## IV. Implementation

We cover the following use cases in this study:

- Device registration and device login.
- Viewing all existing assets in the system.
- Viewing assets owned by the device that is currently logged in.
- Uploading a new device data.
- Simulation of IOT environment using Node Red.

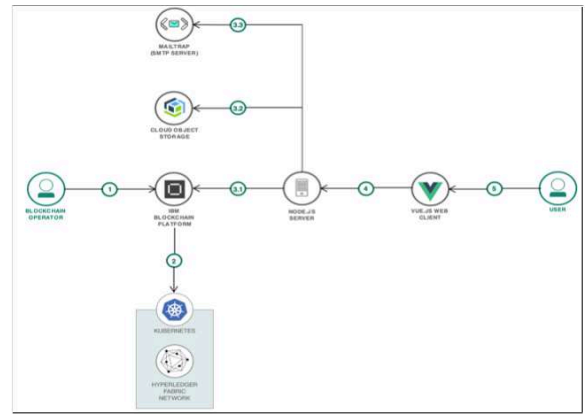

Fig. 7. Application design.

The steps below elaborate Fig. 7:

- The IBM Blockchain Platform service is set up by the Blockchain Operator.

- On an IBM Cloud Kubernetes Service, the IBM Blockchain Platform service constructs a Hyperledger Fabric network, and the Blockchain Operator installs and instantiates the smart contract on the network.

- The Node.js application server creates APIs for a web client by interacting with the deployed network on IBM Blockchain Platform, IBM Cloud Object Storage instance, and Mailtrap Server (fake SMTP testing server) using the Fabric SDK.

- To connect with the network, the Vue.js client uses the Node.js application API.

- The user interacts with the digital asset management application using the Vue.js web interface.

The major components involved are:

- Blockchain platform: provides you complete control over your blockchain network with a user interface that can simplify and accelerate your journey to deploy and manage blockchain components on the IBM Cloud Kubernetes Service.
- The IBM Cloud Kubernetes: Service launches highly available containers in a cluster of compute hosts. A Kubernetes cluster allows you to manage the resources you need to swiftly deploy, update, and grow applications in a secure manner.
- The IBM Blockchain Platform Extension for VS Code: helps users create, test, and deploy smart contracts, including connecting to Hyperledger Fabric settings.
- IBM Cloud Object Storage: is a scalable cloud storage solution with a focus on durability, robustness, and security.
- Mailtrap.io: is a test mail server that allows you to test your email.

Hyperledger Fabric's design is highly flexible and configurable, allowing for creativity, variety, and optimization across a wide range of business use cases, including banking, finance, insurance, healthcare, human resources, supply chain, and even digital music delivery.

Fabric is the first distributed ledger platform that support smart contracts written in general-purpose programming languages like Java, Go, and Node.js rather than domain-specific languages (DSL). The platform is also permissioned, which means that, unlike a public permissionless network, participants are known to one another rather than anonymous and hence completely untrustworthy. While the members may not entirely trust one another (for example, they may be competitors in the same business), a network can be run under a governance model based on the trust that does exist between them, such as a formal agreement or a framework for resolving disputes.

### A. IBM Blockchain Platform

IBM Blockchain Platform is a managed and full-stack blockchain-as-a-service (BaaS) product that lets you install blockchain components in your preferred environment. Clients can use an offering that can be used from development to production to construct, operate, and grow their blockchain networks. It contains the IBM Blockchain Platform console, user interface and VS code extension that makes deploying and managing blockchain components easier and faster. Fig. 8 given below demonstrates the IBM Blockchain Platform components.

*1) IBM Blockchain Platform Console (UI):* This is the user interface for creating and managing blockchain components. You can launch an instance of the IBM Blockchain console and link it to your Kubernetes cluster on IBM Cloud when you provision a service instance in IBM Cloud. Then, on your Kubernetes cluster, you can utilize the console to create and manage your blockchain components. The console is free of charge.

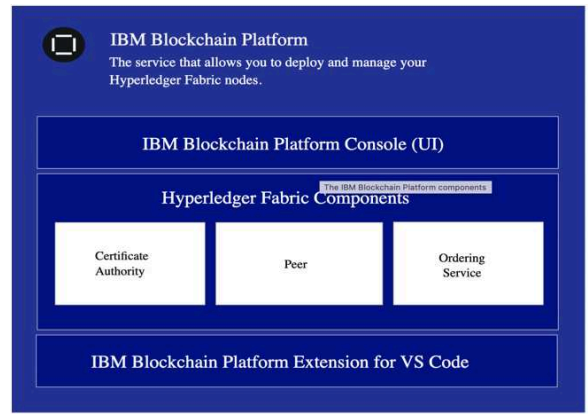

Fig. 8. The IBM blockchain platform components.

*2) Hyperledger fabric components:* The console is used to build and manage Hyperledger Fabric v1.4.12 and v2.2.5 Certificate Authority, peer, and ordering service images, as well as other blockchain components. When these components are deployed into your Kubernetes cluster, the default storage class is used to provision storage for them.

*3) IBM VS Code extension (Development tools):* For writing, packaging, and testing client applications and smart contracts, download the VS Code extension from the VS Code marketplace.

Each CA, peer, and ordering node in the IBM Blockchain Platform requires persistent storage. A pre-configured storage class powered by IBM Cloud File Storage is used when you install a standard Kubernetes cluster in IBM Cloud. You can alter your backing storage, improve your storage class, or use third-party services.

Every cluster on your IBM Cloud Kubernetes cluster has a default storage class that is used to provision persistent storage on IBM Cloud. When you use the IBM Blockchain Platform UI or APIs to deploy a blockchain node to that cluster, the node utilizes this default storage class to dynamically provision the amount of storage you specify on IBM Cloud.

### B. Smart Contracts

A smart contract, or "chaincode" as Fabric defines it, is a trusted distributed application that gets its security or trust from the blockchain and the underlying peer consensus. It is a blockchain application's business logic. Smart contracts have three main characteristics, especially when used on a platform:

- Many smart contracts run in the network at the same time.
- They can be installed dynamically (by anyone in many cases).
- Application code should be treated as untrustworthy, if not malevolent.

The implementation details of the application are based on the the following smart contracts:

- Checking whether the device is registered in the blockchain
- Registered device should only be able to do the login.
- Viewing all existing assets or transactions in the system.
- Viewing assets owned by the device that is currently logged in.
- Uploading a new IOT device data.

We have implemented a smart contract to avoid duplication of assets, duplication of devices during registration, and to ensure that the device is registered before login. Each of these functionalities empower and strengthen the security and confidentiality features offered by the properties of blockchain. First feature ensures that there are no same copies of singular asset. This helps avoid potential complications due to the double-spending problem. Also, makes sure that there is one copy of an asset in the blockchain. If any devices with malicious intent tries to make multiple copies the assets, the device will be denied to do so. Avoiding duplication of devices during registration makes sure that malicious device would not be able to register by using other device's identity.

The device registration smart contract creates a worldwide register, where the unique DID can be stored. The global registry records the device entries so that an audit trail may be created for all events since the device was manufactured. In this proposal, an owner attribute is proposed to allow the device owner to execute operations on the device DID document as well as access to DID methods.

The access control mechanism for these functions is based on the different modifiers specified in this contract. The root of trust is formed with such a level of control over identity formation, and the device owner ultimately has full control over the device identity.

*C. Node-RED*

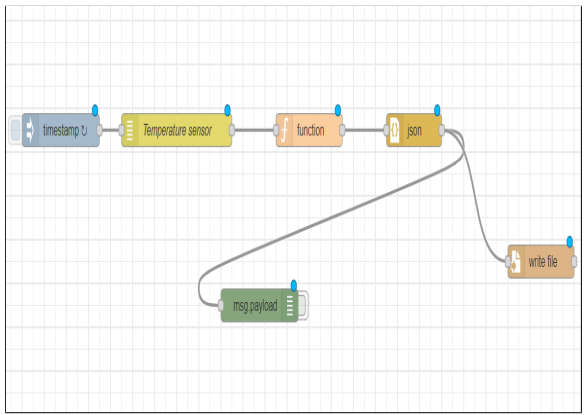

Fig. 9. IoT Device simulation with Node-RED.

Node-RED is an open source visual editor that allows us to connect hardware devices, create flows, and write Javascript functions to control those flows and devices. We used Node-RED to simulate an IoT device, generate data from the device and eventually simulate a network of five devices for the demonstration. Following are the reasons why we chose Node-RED: When it comes to programming with Node-RED, the simplicity is the key parameter. As the name implies, programming is performed intuitively with the use of a small number of actions. The majority of Node-RED tasks may be accomplished nearly entirely through GUI actions. So you can focus on programming, the Node-RED flow editor takes care of creating the application application infrastructure, library sync, integrated development environment. The actual benefit of flow-based and visual programming is high quality. Each component node is a fully functional module that has been thoroughly tested. As a consequence, app writers may concentrate on validating the operation at the join level rather than the information of the node. This is a significant aspect in preventing human mistake at the individual level and ensuring good quality.

```
1  flow.set('random1', msg.payload);
2  msg.payload={'random1':msg.payload};
3  let manufacturer_id = 'ABCDEF00001';
4  msg.payload={'d':{'manufacturer_id':manufacturer_id, 'Time': new Date(), 'temperature':flow.get('random1')}};
5  msg.header={"Content-type":"application/json"}
6  var c = context.get('c') || 0
7  c ++
8  context.set('c',c)
9
10 var path = "~/blockchain/nodered/device1/"+c+".txt"
11 msg.filename = path
12 return msg;
```

Fig. 10. Function node in Node-RED.

For our purpose, initially, we simulated an IoT device with the help of five nodes, as shown in the Fig. 9. First node is the inject node which initializes the flow as per our needs. We have set it to execute the flow every 30 seconds. This means that we will be recording temperature every 30 seconds. Temperature values are generated with help of this node which is a random node. It randomly outputs a value between the 0 to 100. Third node is the function node which receives the value generated by the temperature sensor. In addition to this, a unique DID is assigned to the device which is required for registration and authentication of the device. Also, every time the sensor measures the data, it is stored in a different file. . With the time, these files are generated and act as assets for that particular device. The code shown in Fig. 10 is used for the function to perform the above mentioned tasks. The next three nodes are used to process the JSON string which is created in the function node. This processed string is displayed on the node red debug console with the debug node. Also the data is then written to a file with a "write file" node. We added five such devices in the network to simulate an IoT network. Each of those devices record and store the assets in a defined directory structure. Fig. 11 shows the format in which data is stored in one file. Such file is created every thirty seconds for every five devices.

## V. EVALUATION

The data generated by the network simulated in Node red is used to perform various operations. Two basic functionalities we have provided for the devices are login and register. The devices will use the unique DID assigned to them for registering and then for login. This ensures authentication and adds level of security. An error will be shown if the a device which not registered tries to login. After login (Fig. 12), the device can query its own assets. In addition to this, there is a functionality for the device the device to see the names of assets added by other devices. Fig. 13 shows the format in which data is displayed. The device can see the asset owner, asset name, the timestamp on which asset was added to the blockchain, and a unique data ID which is used to check the duplication of assets in the system.

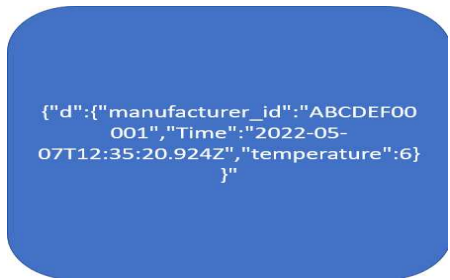

Fig. 11. Data storage in a file.

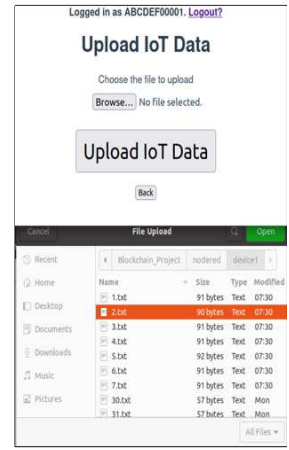

Fig. 14. Feature 4 - Upload new IoT device data (transaction) to the blockchain.

In addition to this, a unique dataID will be displayed which represent the asset that already exists on the blockchain. This helps to prevent possible duplication of data on the blockchain.

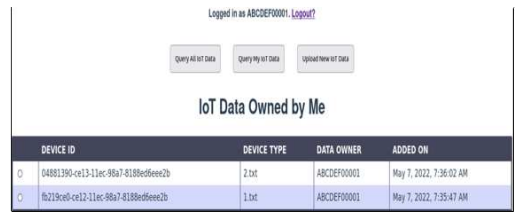

Fig. 15. Feature 5 - Query data owned by the logged in device.

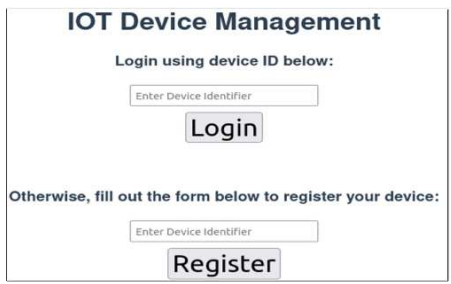

Fig. 12. Feature 1 and 2 - Device login and register.

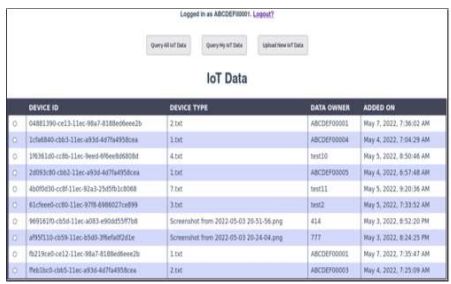

Fig. 13. Feature 3 - Query of all transactions in the blockchain.

A functionality is provided for each device to upload the recorded data to the blockchain. Fig. 14 shows the interface to upload the data. After the data is uploaded, it can be queried in the option to show device's data. Fig. 15 shows how the device's data is shown in the web interface. If device uploads the same data again, an error will be shown which informs the device that the same data or asset already exists in the system.

## VI. RELATED WORK

There have been several inquiries and implementations of Blockchain based identity management solutions, with multiple studies investigating different approaches to achieve a reliable framework. Uport [15] is a self-sovereign identity based identity management system which lays out foundation for better access and authority to the users. It provides a digital description of a user or entity by using Ethereum based smart contracts. This enables Uport to resolve the issue of key loss and recovery. [16] Identifies Uport's key advantages and limitations. It's advantages mainly revolve around the simplicity, ease of use, and the user-centric behavior. However, it is limited by the security concerns related to private key which controls the identity. It also suffers from the restricted size of the smart contracts.

In [17], the authhors present another Ethereum and smart-contract based identity management system which aims to avoid dishonest and harmful certificates issued by certificate authorities that caused major issues in traditional public key cryptography based IDMs. It provides support to multiple operations such as adding and signing attributes where the attributes are building blocks of the identity of the entities such as keys and addresses. However, performance analysis is done only in terms of Gas cost. Also, as this work does not make use of "zero-knowledge" phenomenon, private data possesses risk of security.

In [18], the authors uses Namecoin blockchain which is comparatively more appropriate for storing data than bitcoin to build a privacy focused decentralized PKI system. It provides users authority over what information to reveal and what to keep private. Also, this system allows users to recover previous private keys or identities. However, it is limited by the not restricting other users from altering the keys before the recovery takes place. This results in potential failure of the system if majority of users are corrupted. The study also does not include any performance analysis to compare the proposed system's statistics.

A distinct smart-contract and Ethereum based identity management system is presented in [19]. It focuses on keeping track of reputation of the users apart from other features like privacy and authentication. A token of positive reputation is awarded to the users if they perform actively take part in the blockchain in a positive manner. The study does not compare the system with existing blockchain based identity management systems. Also, real world large-scale data could be used in the experiments to further test the robustness of the system.

Identity management system proposed in [20] provides additional features like voluntary disclosure of identity if something goes wrong with a transaction of that identity. This is achieved with the help of multiple keys and by not associating the transactions done by an entity with each other. Also, in this system, anonymity is achieved by "zero-knowledge" proof protocol. Along with this, identities are also verifiable. The study gives very little insight on how the framework compares and performs with respect to other frameworks. Also, the study does not address how user can keep track of multiple keys.

As discussed above, majority of systems implemented in the studied papers refrain from conducting performance analysis. In [22], the authors outlines existing performance analysis methods for Blockchain technologies. The work analyzes three key tools for performance analysis: BlockBench [23], HyperLedger Caliper [14], and DAGbench [24]. Apart from these tools, the study highlights crucial practices for performance analysis. It also gives importance to having a well organized and documented workload which can help in carrying out the analysis.

The existing solutions for decentralized identity and access management differ from our solution in a few aspects[21]. Firstly, we aim to develop a decentralized identity management using permissioned model of blockchain in contrast to the existing permissionless model. Secondly, we aim to include environmental inputs of devices in order to generate a unique and tamper resistent device identification. Third, we wish to deploy the decentralized identity and access management and the application to track transactions of these devices on the Hyperledger Fabric.

## VII. DISCUSSION

The results obtained from this implementation indicate that decentralized identity systems can significantly improve authentication and asset traceability in IoT environments. Node-RED proved efficient for simulating sensor behavior, while Hyperledger Fabric demonstrated its strength in secure asset registration and querying. Although effective, the model faces scalability challenges and dependency on configuration specifics like consensus mechanisms and cloud environments. Further work is needed to compare performance across platforms and extend usability to real-world IoT networks

## VIII. CONCLUSION

Recent advances in IoT networks promise an ecosystem in which devices can interact and communicate with one another in a peer-to-peer format without the need for a centralized authority to identify and authenticate devices and grant access to offered services. Furthermore, the difficulties posed by the heterogeneity of IoT networks sparked interest in developing decentralized architectures, with a focus on providing decentralized device management functionalities as a foundation that can be extended to provide decentralized services and applications.

This research covers the creation of a decentralized IoT device identification and access management system that can be used to enable peer-to-peer and decentralized IoT applications. We deployed the decentralized identity and access management system to track transactions of the IoT devices on the Hyperledger Fabric using IBM Blockchain platform. This platform offered us the flexibility and ease of deployment by providing Platform Console (UI), Hyperledger Fabric components and Development Tools ( IBM VS Code extension). We created a web-application that allow the devices to register, login, upload and query the IOT data. Finally, we made use of Node-red to simulate a realworld IOT enviroment that allowed us to connect hardware devices and create flows to control those flows and devices.

The following scenarios are addressed in this research:

1) Blockchain registration of a new IoT device
2) Login feature for IoT devices that have been registered
3) Seeing all of the blockchain system's assets and transactions.
4) Seeing assets that belong to the current logged-in device.
5) Adding new data from an IoT device to the blockchain.
6) Smart contract to handle duplication of assets, users, and login without registration.

## IX. FUTURE WORK

The field of decentralized IoT device identification and access management continues to attract interest from academia and business, and it is a hotbed of study and development. Many new issues arise as a result of the growing trend toward decentralized techniques and giving device owners more control over their data. New research venues for expanding this work are to be sought as blockchain technology matures and more IoT corporate use cases arise.